\documentclass{hep99}
\usepackage{epsfig}
\begin{document}

\title{ String/M  Theory }

\author{C P Bachas}
%

\address{Laboratoire de Physique Th{\'e}orique de
 l' Ecole Normale Sup{\'e}rieure\\
24 rue Lhomond, 75231 Paris Cedex 05, France\\[3pt]
E-mails: {\tt bachas@physique.ens.fr, bachas@cpht.polytechnique.fr}}

\abstract{This is a brief review of the present 
status, of some recent developments and 
 of the  open  challenges in  string/M  theory.\dag}

\maketitle

\fntext{\dag}{Plenary talk at the EPS HEP99 Conference, Tampere, Finland}

\section{Why string theory}

  The main  claim of string theory
 is that it  cures the singular
short-distance behaviour of Einstein gravity, without giving up
the fundamental laws of quantum mechanics. Although this is far from
established,  there are serious  indications that the claim could 
indeed be true. In any case no other real  alternatives to string
theory are known at present.  
One  can draw an analogy with spontaneously-broken  gauge theories,
which cured the singular short-distance behaviour of the four-Fermi 
interactions. The resolution of this theoretical inconsistency lead
to the construction of the Standard Model, and one hopes that the 
resolution of the problems associated with 
 quantum gravity  will, likewise, lead us to the
fundamental microscopic  theory that lies beyond.

   A frequent objection is that  while  the Fermi scale was clearly
accessible to  accelerator experiments,
quantum  gravity need not manifest itself below $M_{\rm Planck}\simeq
1,2\times  10^{19} GeV$ ,  a  scale surely out of direct experimental
reach. 
String theory  predicts, however,  many types of new physics:
supersymmetric partners, Kaluza-Klein and string excitations, enhanced
gauge symmetry,  and a larger gravitational sector. All of these are
required   for consistency, and  the scales at which they should appear 
is an  important open question  to which
I will return later in this  talk. 
Even if one adopts the conservative hypothesis that non-gravitational
physics much below $M_{\rm Planck}$ can be described by a minimal
supersymmetric extension of the Standard Model (MSSM), 
it is  increasingly hard to discard string theory as irrelevant.
First, the MSSM is an effective theory with so many arbitrary parameters
that it has little predicitive power without extra assumptions. 
Second, physics at a superheavy scale can have many indirect 
manifestations: we are used to associate
 neutrino masses, superlight axions, proton decay or primordial cosmology
with the physics of conventional grand unification, and
 $M_{\rm GUT}\simeq 2\times 10^{16} GeV$ is at most one order
 of magnitude below the fundamental string scale.

   The search for a unified theory has been the main preoccupation of
string theorists over the past fifteen years. Effective  string excitations
appear however also in many  other contexts, and most notably in the low-energy
limit of confining gauge theories like QCD. 
The early development of dual resonnance
models was in fact  triggered by 
the  observed Regge behavior of meson  scattering, and the hope to
find a controllable string approximation to QCD has been an important
motivation  in the subject ever since. The idea has been   revived recently
in a beautiful and unexpected way that I will discuss.
Though it is too early to see where it will lead, it does illustrate how
string theory has replaced quantum field theory as the new exciting
theoretical frontier. 

 The plan of my talk is as follows: I will first briefly  review the status
of string theory up to the early 90's, and the rapid subsequent 
developments associated with duality symmetries, black holes and  D-branes.
 I will then focus
on three issues that have received attention over the past year:
efforts to do away with  supersymmetry, the AdS/CFT correspondence, and
the question of string theory scales or of whether we  live  on
a brane. This list is by no means exhaustive,  it reflects partly
my taste and partly my ignorance. There are some interesting topics
I will not touch -- cosmology or non-commutative theories to name a
couple -- and many more interesting  papers I do not cite. 
Time  limitation and ignorance is more frequently at cause
here than taste.

\section{Perturbative string theories}

   There are five known perturbative string theories, or rather
consistent sets of Feynman rules (reviewed in many books
on the subject \cite{stringbooks}): 
 the type IIA and IIB theories, the
heterotic theories with gauge group  $SO(32)$ or $E_8\times E_8$, and
the  type I theory with gauge group $SO(32)$. Except for the type I theory
whose quanta are unoriented closed and open strings, all the others
have only oriented closed strings in their spectrum. They are all
defined in ten space-time dimensions, and have N  unbroken
supersymmetries, where  N=2 for the type II theories,
 and N=1 for the rest. 
One of their  key ingredients  is  anomaly cancellation. Anomalies
are quantum failures of classical symmetries, which  are fatal 
when the symmetries in question are local.
Anomaly cancellation
 requires for instance  complete families of quarks and leptons
 in the Standard Model.
The effective N=1 supergravity plus super Yang-Mills
theories in ten dimensions  
were believed to be plagued by both gauge and gravitational  anomalies
 \cite{AW}. 
It was the important  discovery of Green and Schwarz \cite{GS}
to show that string theory avoids the problem in a subtle way:  
the symmetry is
violated classically, and  restaured by quantum corrections as
illustrated schematically  in figure 1.

\begin{figure}
\begin{center}
\leavevmode
\epsfxsize=6.5cm
\epsfbox{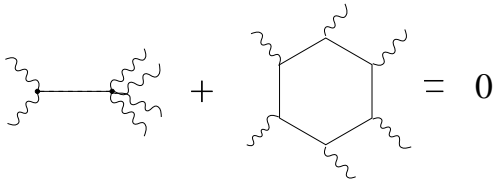}
\end{center}
\caption{The Green-Schwarz anomaly-cancellation 
between a tree and a one-loop hexagon diagram.}
\end{figure}

 Having found perturbatively consistent theories of gravity, the
next  step  was to see if they  could be made
to fit our low-energy world. To do that  one had to compactify six
of the ten space-time dimensions, obtain  a gauge group large enough to
contain  ${\rm SU(3)}_c\times {\rm SU(2)}\times {\rm U(1)}$,
 and generate  on the way 
chiral matter in which the quark and lepton families could fit.
Now N=1,2 in ten dimensions corresponds
to N=4,8  in four, and 
extended supersymmetry does not allow chirality in four
 dimensions (4d).
 In the process of compactification one should 
thus break either all, or all but one, 
of the original supersymmetries. 
For reasons that will  become clear 
later  this  second alternative has been  privileged, so that 
a lot of  effort went into constructing 
`semirealistic' N=1 compactifications to 4d. The two  type II theories
proved  too restrictive  to accomodate the Standard Model \cite{DKV},  
the type I theory was at the time technically more awkward to deal with,
but for the  heterotic theories many such compactifications
were  found and extensively analyzed in the mid eighties  \cite{4d}.

   The main challenge of string theory still today  is  related to this
embarassement of riches. It is the question  of 
{\it vacuum selection and stability}. In a nutshell,  if a 
supersymmetric vacuum is a good first approximation to our world,
there exist a  zillion candidates for the job, and if not there are 
no stable candidates whatsoever. Part of the huge degeneracy of
supersymmetric vacua is due to the existence of massless moduli. These
are parameters of the compactification,  which
 appear as 4d scalar fields with
vanishing potential to all orders in the string coupling constant.
Their expectation values are  
 thus perturbatively undetermined -- a nuisance since the masses, 
couplings and other properties of the effective theory 
change  drastically as one moves around  moduli space. 
Supersymmetry breaking, possibly triggered by non-perturbative gaugino
condensation \cite{gaugino}, could in principle lift this large
degeneracy but it has proven hard to do  reliable calculations  in
this context. Things are even worse for non-supersymmetric 
compactifications, where perturbation theory itself breaks down.
Finding  a stable non-supersymmetric
vacuum in a theory containing  gravity is
of course a very ambitious task, since it amounts to
solving among other things the  
problem of the  cosmological constant.


\section{Dualities and M theory}

 It is very likely that the solution to these long-standing problems
will require  a definition of string theory that goes beyond the
Feynman-rules of the previous section. 
We are still lacking such a definition today,
but  progress towards this goal has  been made.
A key idea, which  has opened a  window into the non-perturbative
 structure of the theory,  is the idea of duality symmetries. 
It can be illustrated with the Georgi-Glashow model,
in which a  SO(3) gauge group breaks  down 
 to a U(1) by the vacuum expectation
value ($v$) of a higgs-triplet field.\footnote{This was an early candidate
for  a theory of the electroweak interactions, which failed to account
for   weak neutral currents.}
At weak gauge coupling ($g\ll 1$) the spectrum of this theory  
contains, in addition to a neutral higgs, three other 
stable particles:
\begin{trivlist}
\item \hskip 0.5cm -- a photon ($\gamma$)
\item \hskip 0.5cm --  a charged gauge boson (W),  and
\item \hskip 0.5cm --  a magnetic monopole (M).
\end{trivlist}
The latter arises as a soliton (discovered by `t Hooft and Polyakov) 
and solitons are  much heavier than elementary quanta,
\begin{equation}
m_W\sim gv \ \ll  \ \ m_M\sim v/g\ .
\label{masses}
\end{equation}
The above inequality is valid at weak coupling, but if we allow
$g\gg 1$
the roles of the electric and magnetic charges are naively exchanged.
 It is thus
tempting to suppose  that there exists a `dual' perturbative 
 formulation
of the strongly-coupled  theory, 
 in which  monopoles are the   fundamental quanta 
(coupling with strength $\tilde g \sim 1/g$), and 
the  W-bosons arise as  solitonic excitations.

\begin{figure}
\begin{center}
\leavevmode
\epsfxsize=7cm
\epsfbox{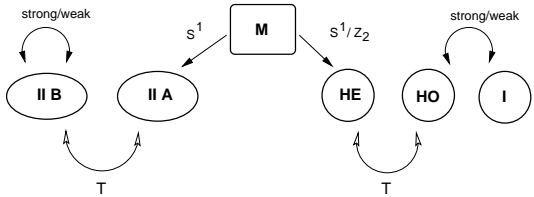}
\end{center}
\caption{The web of string theory dualities.}
\end{figure}

The problem with this kind of conjecture
 is that most of the time  there is nothing  one
can calculate  to test it. The  mass formulae
(\ref{masses}) for example  receive in general
both perturbative and non-perturbative corrections, 
so at strong coupling they need not be even qualitatively correct.
 There is however one  notable exception: extended
N=4 supersymmetry protects both the mass and the gauge coupling 
from quantum corrections,
and  the idea could  in this case be put to the test
with success \cite{MOWS}.  As shown by Seiberg, Witten and others,
electric/magnetic
duality is in fact also a powerful tool in 
less supersymmetric,  N=2 and N=1 
gauge theories,  but describing  these celebrated results
is  outside the scope of the present talk (for  reviews  see
for example \cite{SW}).

  Dualities have their natural habitat in string theory
\cite{dual} (for precursor ideas see \cite{prec}, for reviews
 see \cite{dualrev}). They take many different forms:
coupling constant inversion (S duality), inversion of the compactification
volume in string units (T duality), or coupling/volume interchange
(U duality). The five perturbative string
theories of the previous section  are related by such a web of dualities,
as illustrated in figure 2. They are thus believed to 
be dual descriptions of the same physics, each  best
 adapted in different regions of the moduli space of some fundamental
underlying theory named ${ M}$. Though we  do not have a precise
definition of M theory, one striking fact about it is that it
exists in eleven flat space-time dimensions,  where its low-energy
limit coincides  with the maximal supergravity of Cremmer, Julia and
Scherk \cite{11d}. Compactification of  the eleventh dimension on a circle
\cite{dual} or  a line segment  \cite{HW} leads  to the type IIA
or heterotic $E_8\times E_8$ theories, at weak coupling when the circle
or segment are very small.

 Dualities have transformed our
 thinking about string theory in many ways.
First they represent an extraordinary unification of principle:
the theory of quantum gravity seems to be unique (there aren' t five
different consistent theories), and even if
vacuum selection and stability remains  a puzzle, it is
conceptually comforting  that the issue is at least 
 entirely dynamical. Second,  they have uncovered  previously
inaccessible regions of moduli space, with new exciting 
phenomenological possibilities. Third, they have given us new tools for
understanding the still mysterious physics of black holes.
Finally, duality multiplets 
include besides fundamental strings a new class of excitations,
 whose role has been pervasive in the subject ever since: Dirichlet
and other p-branes.

\section{D-branes and black holes}

   A p-brane is an excitation with p-dimensional spatial extent:
p=0 for particles, p=1 for strings, p=2 for membranes and so on
(the list would stop there if we did not live in higher dimensions).
Many conventional field theories
 have p-brane solitonic excitations.
These are solutions of the classical non-linear field equations,
localised in the transverse  dimensions, and  characterised by finite
tension (mass/volume) and charge densities. 
The  `t Hooft Polyakov monopole is such a soliton, 
other  examples  are the cosmic
strings or domain walls of conventional  grand unified models. 
It was the
important insight of Polchinski \cite{Dp}
 (for reviews see \cite{Dprev})
to recognize that  type II string theories contain a particular
class of such excitations that
admit a very simple description as D(irichlet) branes. 
In a theory with only closed strings in the bulk,
the   (p+1)-dimensional worldvolume 
of a Dp-brane is a space-time defect on which fundamental open
strings can  end.

  It turns out that this almost poetic definition determines 
the properties of the soliton precisely, with no need to ever solve
any non-linear field equations explicitly.\footnote{The technical reason is
that the classical string-field equations are equivalent
to the requirement  of
(super)conformal invariance on the worldsheet, which is  respected
by the Dirichlet boundary conditions of  string endpoints.}
Consider for example the interaction of two D-branes. To leading order
this is given by the exchange of a closed string (as in figure 3), 
whose massless  modes are the graviton and its supersymmetric cohort. 
Only these contribute at large brane separation, but to evaluate the
interaction we need to know how they couple to the solitons -- we
need to know in particular the D-brane  charges
 and their tensions. The same
diagram has however also a dual interpretation: it is a Casimir
force due to the vacuum fluctuations of open  strings, much
like the Casimir force between two ordinary superconducting plates.
This only depends on the spectrum of stretched open strings, so
that the result  can   be evaluated readily \cite{Dp}\cite{move}
both for  large brane separations and also  at short distances, 
where the supergravity approximation is not valid.

\begin{figure}
\begin{center}
\leavevmode
\epsfxsize=4.5cm
\epsfbox{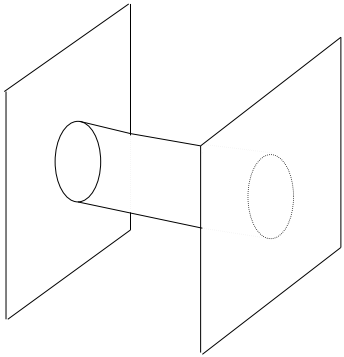}
\end{center}
\caption{Leading interaction of two D-branes through the exchange
of a closed string. The diagram has the  dual interpretation 
as  Casimir force -- see  text.}
\end{figure}

   The importance of D-branes derives from two basic facts:
(a) they can `trap' in their  worldvolume non-abelian gauge fields,
in addition to spin 1/2 and spin 0 excitations,\cite{witt}
and (b) being solitons in a theory of gravity they can be assimilated 
in  appropriate circumstances to  large  
semiclassical black holes \cite{black}. 
The first of these facts  is illustrated in figure 4: the open strings living
on a single D-brane reduce at low energies to an abelian
 gauge boson plus  its
supersymmetry partners. Putting $n$ D-branes together gives rise
to  $n^2$ different types of open strings, whose  low-energy
excitations correspond now  to  a supersymmetric U($n$) gauge theory.
Separating the branes breaks this gauge symmetry spontaneously
to ${\rm U}(1)^n$, while assembling different types of D-branes
can reduce the number of unbroken supersymmetries. 
 When $n$ is large such  collections  of branes become
heavier and heavier, and 
should  eventually behave like large 
 semiclassical black branes  or black holes.

\begin{figure}
\begin{center}
\leavevmode
\epsfxsize=4.5cm
\epsfbox{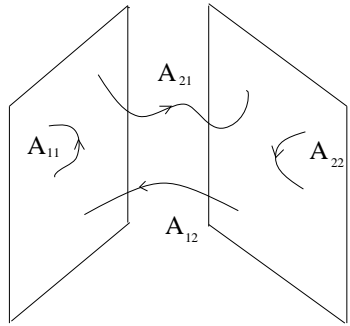}
\end{center}
\caption{The four different types of open strings in the presence of
two D-branes. Their low-energy limit corresponds to  
a U(2) gauge theory.}
\end{figure}

   The super Yang-Mills/supergravity descriptions are a priori 
mutually exclusive, much like the two dual descriptions
of the Georgi-Glashow model.  
The argument for this goes roughly   as follows: the mass of the would-be 
black hole in units of the fundamental string length
is  $M_{\rm BH}\sim  n/g_s$,  where $g_s$ is the 
string coupling constant  and $\sim 1/g_s$ is the tension of a 
D-brane. The coupling of the effective super Yang Mills theory is
$g_{\rm YM}^2 \sim g_s$, while 
Newton's constant  is $G_N \sim g_s^2$ -- this is
because in the topological expansion a closed-string loop has
double the  weight of  an open loop. Now for the supergravity
description to be valid the Schwarzchild radius of the  black hole 
 must be much bigger than the  string length,
\begin{equation}
M_{\rm BH}G_N  \sim  ng_s \gg 1 \ \ \ {\rm (supergravity)}
\end{equation}
(a more careful argument should take into account
compactification radii). 
The Yang Mills description, on the other hand, is  valid 
when the loop counting
parameter is  small
\begin{equation}
ng_{\rm YM}^2  \sim  ng_s \ll  1 \ \ \ {\rm (Yang-Mills)}
\end{equation}
where $n$ is the number of gauge bosons running in a  loop.
The two descriptions have thus non-overlapping ranges of validity.

Fortunately, supersymmetry  can come once more
  to our  rescue. For certain
extremal black holes corresponding to supersymmetric 
configurations of branes, non-renormalization allows
the extrapolation of some results
 from one limit to the other.  This is true in particular
for the entropy which can be calculated in the field theory
at weak coupling,  and then compared to the 
Beckenstein-Hawking area law \cite{BH} 
obtained by thermodynamic arguments
on the black hole side. The agreement of the  two results 
\cite{black}(for a detailed account see
\cite{blackrev}) gave the first   microscopic derivation of
black hole entropy. This  has boosted our confidence that string
theory is a consistent microscopic theory of gravity
even if, to be sure, the puzzles of black hole evaporation remain.

\section{Is there life  without supersymmetry?}

   Most of the discussion up to now has relied heavily on 
supersymmetry: duality checks, the microscopic
black hole entropy calculation and vacuum stability, 
 all depend crucially   on it. Any sharp statement about non supersymmetric
situations is thus a priori interesting, and some progress in
this direction has been made over the past year, thanks mainly
to the work of Sen (for a review and references see \cite{nonBPS}). 
His idea was to consider non supersymmetric states 
(these are states in long or non-BPS multiplets 
of the supersymmetry algebra)
which are protected from
decaying by some discrete symmetry. The simplest  example is provided by
heterotic string states in spinor representations of SO(32).
The lightest such state must be stable even at
 strong coupling,  and if duality is correct this state  should
 exist in the dual type I description of the same theory. Sen identified
this  state  as a controlled remnant in  the annihilation of a
D-brane anti D-brane pair, and succeeded to calculate  its  mass
and other properties.  Many more examples have since been found,
and classified by the elegant mathematical theory called
K-theory \cite{Kth} (this is a topological classification, it has
no claim to being a fundamental theory of nature like M !)

   Can this help us to construct non-supersymmetric stable vacua?
 Despite some brave recent attempts in this direction 
\cite{zeroco}  I dont think the answer is anywhere in sight yet.
Let me recall what the situation was in older attempts to break
supersymmetry in the heterotic string. The breaking scale 
could be (i) the heterotic string scale $M_h$ \cite{o16}, 
(ii) some Kaluza-Klein
scale $1/r$,\cite{SS} 
 or (iii) some dynamical scale $\sim M_h {\rm exp}(-1/g_h^2)$, 
if the breaking was induced by gaugino condensation 
\cite{gaugino}(here $g_h$ is
the heterotic string coupling).
 The fact that only the first two
options are perturbative,   motivated some early speculation that
string theory should have TeV-size radii \cite{ABLT}. 
Weak coupling forces however, as I will explain in section 7, 
both $M_h$ and $1/r$ to be near the Planck scale, 
so only the third possibility
looks theoretically plausible. If one gives up low-energy supersymmetry
altogether, vacuum instabilities at the string scale  render
perturbative calculations  unreliable -- even if the problem
can  sometimes be postponed \cite{zeroco}. In the gaugino condensation
scenario things  `look better' but the important puzzles remain: 
why doesn't  vacuum energy at the breaking scale
drive the modulus $g_h$ to zero, and why does the energy vanish at the
minimum anyway?

   The type I theory does seem to offer some novel possibilities.
The vacua of this theory contain generally  D-branes,  as
well as another type of defect, or rather `end of the world',  called
orientifold planes \cite{augusto} (for reviews see
for example \cite{Irev}). These should fill all  three 
non-compact   dimensions of our universe, or else Poincare  invariance
would be lost, but they may  be localized in the compact space.
Supersymmetry can be broken on these branes and walls,  while being
to leading order preserved in the ten-dimensional bulk \cite{mag}
(a similar phenomenon has been
 discussed in the Horava-Witten limit  of the
strongly-coupled heterotic theory \cite{Ho}). 
Can we then use the  stable non-BPS branes discussed  above to construct
a non-supersymmetric type I vacuum? The difficulty is that such branes
carry in general energy and, as a result,  gravitate. Since in a compact space 
`gravitational  flux' has nowhere to escape to, 
the total energy should vanish. This  is however precisely our old
friend the cosmological constant problem.

\section{New hope for the QCD string ?}

I have already mentioned in the introduction
that the Regge behaviour of meson scattering was
the earliest experimental manifestation of relativistic string theory,
and the first motivation for developing dual
 resonance models. 
QCD has taken much of the steam out of this program, but has not lead
to a  quantitative control of  strong interactions except at
high energies where the theory is asymptotically free.
Effective chiral lagrangians
 and lattice gauge theory provide useful
approximations at lower  energies, but the very property  of  
confinement is a strong hint   that a better
 string-theoretic description may  exist.
 Past attempts 
to use the known string theories for QCD
have failed in many respects: 
\begin{trivlist}
\item \hskip 0.5cm -- the critical dimension is 10 (or 26)
\item \hskip 0.5cm --  there is a massless spin-2 particle 
\item \hskip 0.5cm --  what about parton behaviour?
\end{trivlist}
In short, the known flat-space string theories  have only a
very remote resemblance to real QCD.

  In a separate  development
`t Hooft pointed out \cite{largeN} that Yang Mills
 theory retains its essential properties (confinement and
asymptotic freedom) but simplifies considerably  if one
takes  the number of colors $n$ to be large. Keeping the
combination 
\begin{equation}
\lambda = n  g_{\rm YM}^2
\end{equation} 
fixed in the limit, he showed that  Feynman diagrams admit
a topological expansion reminiscent of the loop expansion of
string theory.  Diagrams that can be drawn
on a plane are weighted by $n^2$,
those that can  be drawn on a torus by $n^0$,  and so on down the
line.  Of course $n=3$ in real life, but this could still be a nice
expansion with some luck!  
Its leading term in which only  the planar diagrams 
survive  has, in particular,  a `classical smell' in it. 
The search for this classical limit (or `master field') has been
pursued over the years with little success. More recently, coming
from different considerations, Polyakov proposed \cite{polya}
that the QCD string must live in a 
curved geometry with more than four space-time dimensions.

  These ideas have been revived and made very concrete in the last
couple of years 
\cite{kleb}\cite{malda}\cite{holo}(a complete
 review with an extensive
list of references is \cite{revAdS}).
 Progress came from a closer study of the
 D-brane/black hole correspondence of the previous section, and
was crystallized into a sharp conjecture by Maldacena \cite{malda}.
 His argument
was to compare the following two different limits of the same
system:\hfil\break
(i) in the D-brane description  the limit where  the fundamental
string tension goes to infinity. What survives is a gauge field theory
on the branes decoupled from the gravity in the bulk, schematically
\begin{equation}
 ({\rm brane\ YM})\oplus ({\rm bulk\ gravity})
\label{lim1}
\end{equation}
Gravity  decouples  because Newton's constant has dimensions of inverse
string tension to some positive power. \hfil\break
 (ii) from the point of view of a distant observer, the limit 
in which his calorimeter  can only measure vanishingly small energies.
What our observer  sees are very  long-wavelength gravitons  in the bulk,
decoupled from {\it all} possible excitations near the horizon
where they are redshifted to zero energy, 
\begin{equation}
 ({\rm string\ near\ horizon })\oplus
 ({\rm bulk\ gravity})
\label{lim2}
\end{equation}
Equating (\ref{lim1}) and (\ref{lim2})
leads  to the conjecture  that the field theory on the brane has a
dual description as a string theory in the near-horizon geometry 
of a black hole. 

  This correspondence is often called AdS/CFT because in the 
best-studied examples  the field theory on the brane is conformal and
the near horizon geometry is  anti de Sitter.
In the special case of n D3 branes the  field theory is the N=4
supersymmetric version of SU(n)  Yang Mills, whose parameters
are identified on the string side as follows
\begin{equation}
 \lambda \sim (R/l_s)^4       \ \ {\rm and} \ \ \ n  \sim  (R/l_P)^4
\end{equation}
Here $R$ is the radius of the AdS geometry, $l_s$ the
string length scale and $l_P$ the ten-dimensional Planck length.
This identification renders clear  the meaning of the `t Hooft expansion:
 $1/n$ corrections are string loop effects and as 
$n\to\infty$ the theory becomes indeed classical -- but this is a
classical string theory in a curved higher-dimensional space-time!
Now classical string theory is described by some conformal invariant
2d model -- so you may think we are done --  but the model
corresponding to this particular background has proven  hard
to deal with \cite{RRback}.
 Definite statements can thus be made only in the
 supergravity limit where one may ignore stringy effects.
This amounts to taking  $\lambda\to\infty$,
which is strong coupling on the gauge theory  side.


Now isn't strong coupling all we are after in real QCD?
Unfortunately not,  the reason being very  roughly as follows: 
in real QCD the coupling runs with energy, which would  translate  
into  varying curvatures  as one moves radially in the near horizon
geometry on the string side (for explicit realizations see
\cite{qcd1}\cite{qcd2}\cite{klebtse}). 
 Supergravity allows us to only control the
region in which curvatures stay small, the rest has to be excised away.
This corresponds to imposing an ultraviolet 
cutoff in the gauge theory, so what 
we really have  is a cutoff theory at strong bare coupling.
To go to the continuum limit we must
scale the bare coupling to zero as we take the
 cutoff to infinity -- but this
drives us precisely to the region that supergravity 
does not control. We  face therefore
 the same problem as  lattice gauge theory, 
but transformed in a surprising
way: to approach the continuum limit we need non-perturbative control
over a two-dimensional
 $\sigma$-model. Whether this model will prove in the future tractable
 remains to be seen, but the ideas behind this surprising duality are
 most likely here to stay.


\section{The case for traditional  unification}

 Let me come finally  to the question that is of more direct
interest to the audience here: `what are the scales of string
unification?'  or put more provocatively, 
`will we see strings and extra dimensions at the LHC?'
(this is also discussed in Peskin's talk \cite{peskin})
The conventional (and conservative) hypothesis  
is that the string, compactification and Planck
scales lie  all to within two or three orders 
of magnitude from each other, and are hence far 
beyond direct experimental reach. The non-gravitational 
physics at lower energies is thus  
described by a 4d  supersymmetric
quantum field theory (SQFT), which  must at least include  
in it  the MSSM. This conventional hypothesis is supported 
 by the following three solid
facts~:  (i) Softly broken SQFTs
 {\it can}  indeed  be extrapolated consistently to near-Planckian
energies without destabilizing the electroweak  scale~;
(ii)  the hypothesis is (almost)  automatic in
the weakly-coupled heterotic string theory, and 
(iii) the minimal (or `desert') string-unification
 assumption  is in remarkable  
agreement with some of the measured low-energy  parameters of our
world.

  The first fact is the well-known reason for introducing supersymmetry
in particle physics  in the first place. 
Softly-broken supersymmetry  solves the 
technical aspect of the gauge-hierarchy problem by preventing
radiative corrections from  driving   $M_{\rm weak}$
to the ultraviolet cutoff. I will return to this point later.

The second argument is  less known to non-string theorists and 
 goes  as follows:  in the weakly-coupled heterotic    
theory both  the graviton and the  gauge  bosons
are massless modes of a  closed string. They thus
live  in  ten-dimensional spacetime and
interact  at tree-level through the sphere diagram.
 The effective four-dimensional  Yang-Mills and
Einstein actions, obtained by dimensional reduction,  therefore read
\begin{equation}
{\cal L}_{\rm gauge} \sim {(rM_h)^6\over g_h^2}\; {\rm tr} F^2 \ \ 
\end{equation}
and 
\begin{equation} 
{\cal L}_{\rm grav} \sim  {r^6 M_h^8\over g_h^2}\; {\cal R}\  ,
\end{equation}
where  $M_h$ and $g_h$ are the heterotic string  scale and string
coupling constant, while $r$ is the typical compactification 
radius. From the coefficients of these actions  one  can read the 
four-dimensional  gauge coupling and Planck mass, 
\begin{equation}
 g_{\rm YM} \sim { g_h \over (r M_h)^3}\ , 
\label{eq:gin}
\end{equation}
and 
\begin{equation}
M_h  \sim \; g_{\rm YM}\;  M_{\rm Planck}  \ .
\label{eq:gins}
\end{equation} 
Factors of $2$'s and $\pi$'s in these relations
are irrelevant to  our
arguments and have been dropped. 
Suppose now that $g_{\rm YM}$ is of order one, or 1/10 it does
not really matter. Then the
universal relation (\ref{eq:gins}) tells us that the
string scale is tied automatically  to the Planck mass \cite{G}.
Furthermore  in view
of relation (\ref{eq:gin}) we must have  $rM_h$ also of order
one, or else the string theory will be strongly coupled.
\footnote{T-duality tells us that the radii canot
 be smaller than the string
length.} Thus all scales are tied up together and there is little leeway
for abandoning the traditional hypothesis within the weakly-coupled
heterotic theory.

  Let me come finally to the third argument in favour of  the
 traditional hypothesis. It is a well-known 
 observation \cite{GQW,SU} (for more recent discussions see \cite{LP})
  that if we  extrapolate the three measured low-energy 
gauge couplings with  the
$\beta$-funcions of the MSSM,  they meet at a scale $M_U\simeq  2\times 
10^{16}$ GeV. This is consistent with the one-loop formulae
\begin{equation}
{\alpha_i}^{-1}(\mu ) = {\alpha_U}^{-1} + b_i \; {\rm log} {\mu \over M_U}
+ \Delta_i
\label{eq:unif}
\end{equation}
where $b_i$ are the $\beta$-function coefficients, and 
$\alpha_U$ is the fine structure constant at $M_U$. 
The threshold corrections 
$\Delta_i$ parametrize  our
ignorance of  the details
of the theory at the unification scale,  and of the details of
supersymmetry breaking. 
Assuming that the  $\Delta_i$ are negligible, 
 and treating   $M_U$ and
$\alpha_U$ as input parameters, we have one
prediction which is verified by LEP data  at the level of a few
percent. Note that it is the existence of the energy desert
which  renders the above
  prediction  meaningful and robust. 
Threshold effects make  a few-percent correction to differences 
of couplings only because  the logarithm  in equations
 (\ref{eq:unif}) is very large.

\begin{figure}
\begin{center}
\leavevmode
\epsfxsize=6.5cm
\epsfbox{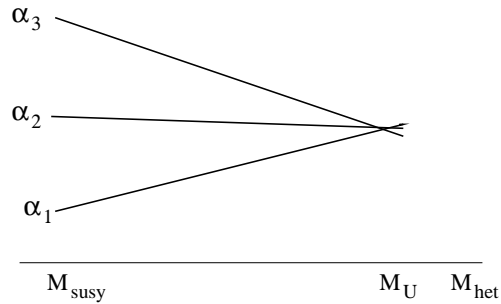}
\end{center}
\caption{Schematic drawing of gauge coupling unification in the MSSM.
On a logarithmic scale the discrepancy between the `experimental' and
the string-theory  unification scale is a few percent.}
\end{figure}

   Gauge coupling unification is an indirect hint for the existence
of the heavy scale $M_U$, 
and also for low-energy supersymmetry,  since 
it works much better with the MSSM $\beta$-functions than with those of
the SM. It is also a  hint in favour of the traditional
string unification, because  $M_U$ is remarkably close --
 though not exactly equal -- to $M_h$. Since this latter is related to
the experimentally known Planck mass and to $\alpha_U$, it is
in principle calculable. 
A careful calculation \cite{Ka}  gives
$M_h \simeq 5\times 10^{17}$ GeV. On the appropriate logarithmic scale
this differs from the `experimental value' $M_U$ by a few percent,
a rather  remarkable a priori agreement\footnote{
For a detailed discussion of
 the  small discrepancy see \cite{Die}}
 illustrated schematically in figure
5. After all there was  no a priori reason why the extrapolated low-energy
couplings should not have met, if at all, 
 at say $10^{35}$ GeV~!


\section{Do we live on a brane?}

  I have already mentioned that
dualities have uncovered some
previously inaccessible regions of the moduli space of M-theory.
The description of the vacuum in these regions involves various
branes that can trap non-abelian gauge fields in their worldvolume:
D-branes, and their dual 
heterotic fivebranes \cite{Wz} and  Horava-Witten walls \cite{HW}. 
Such  vacua could  realize the old idea 
\cite{RS} according to which 
we might be  living  on a  brane, which traps
in it the Standard Model fields.\footnote{A more extreme idea
is that gravity itself could be trapped on a brane\cite{RSun}.}
 Since gauge fields and
gravity live now in different spaces, the universal relation
tying up string, compactification 
and Planck scales is  generally  lost.

Let me illustrate this in  type I theory, whose compactifications
involve  as we saw 
collections of D-branes and orientifolds. 
The graviton (a closed-string state) 
lives  in the ten-dimensional bulk, and interacts to
 leading order through the sphere diagram. The open-string gauge bosons
on the other hand are  localized on the worldvolume of some
collection of branes (I will call them our brane world)
 and interact via the disk diagram, 
 which is higher order in the topological expansion.
The Yang Mills and Einstein actions in four dimensions read
\begin{equation}
{\cal L}_{\rm gauge} \sim {(rM_I)^{6-n}\over g_I}\; {\rm tr} F^2 \ \ 
\end{equation}
and 
\begin{equation} 
{\cal L}_{\rm grav} \sim  {r^{6-n}{\tilde r}^n
 M_I^8\over g_I^2}\; {\cal R}\  ,
\end{equation}
where $r$ is the typical radius
 of the $n$  compact dimensions transverse to our
brane world,   $\tilde r$ the typical radius
 of the remaining (6-$n$) compact longitudinal
dimensions, $M_{\rm I}$ the type-I string scale and $g_{\rm I}$
the string coupling constant. By  T-dualities we can 
 ensure  that both $r$ and $\tilde r$ are
greater than or equal to the fundamental string length.

 Unlike the case of the heterotic string
there is no universal relation between $M_{\rm Planck}^2$ 
(the  coefficient
of the Einstein term), $M_I$ and $g_{\rm YM}^2$ (the inverse coefficient
of the Yang Mills term). The 
 radius  $r$ of the transverse space, in particular, enters only 
in the gravitational action, so by taking it to be large we
can make the gravitational interaction  weaker and weaker, while
keeping  $M_I$, $g_I$ and  $g_{\rm YM}$ fixed. 
This is explained intuitively by  figure 6. 
Thus type I string theory is much 
more flexible (and  less predictive)  than its
heterotic counterpart. 
This added flexibility can be used to `remove'  the
order-of magnitude discrepancy
between the apparent unification and string scales
\cite{W}.
If one decides however to ignore gauge coupling
 unification (efforts to reconcile matters are made in \cite{DDG})
 as accidental 
there is no reason why not to  lower  $M_I$  further,   to
an intermediate \cite{Kar} or even to
 the  TeV scale \cite{ADD}\cite{Ly}.
Keeping for instance $g_{\rm I}$ and 
 $({\tilde r} M_{\rm I})$ fixed and of order one, 
leads to the condition
\begin{equation}
r^n \sim {M_{\rm Planck}^2/ M_{\rm I}^{2+n}} .
\label{eq:mm}
\end{equation}
A TeV string scale would then require from  $n=2$  millimetric 
to  $n=6$  fermi-size dimensions transverse to our brane world.

\begin{figure}
\begin{center}
\leavevmode
\epsfxsize=5cm
\epsfbox{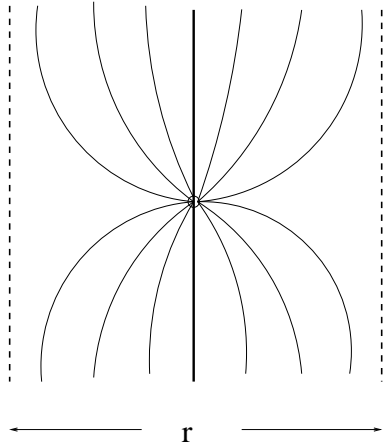}
\end{center}
\caption{The spreading of flux in the transverse compact dimensions
that could  be responsible for the weakness of gravitational forces
in a brane world.}
\end{figure}

 What  focused  attention \cite{ADD}  to the brane-world idea
 was (i) the realization 
that it cannot be generally  ruled out by  existing data,  even in
 the most  extreme case of `TeV-ish'  string scale and 
millimmeter-size  transverse space, and (ii) the hope
that this extreme case could be a new solution to the problem of
the gauge hierarchy. Let me start by  explaining  the first point.
 Gravity is hard to test  at submillimeter  distances because of the
 large  background of residual
 electromagnetic interactions. The ratio for instance of the  Van der
 Waals to Newtonian force 
 between two hydrogen atoms a distance $d$ apart is 
\begin{equation}
{F_{\rm VdW}\over F_{\rm grav}} \sim 
\left( {1 { mm}\over d} \right)^5 \ .
\end{equation}
At    $d= 10\mu m$ Newton's force is thus ten orders of magnitude
weaker than Van der Waals.   
As a result the present-day data  \cite{gravi} allows  practically any
modification of Newton's law,  as long as it is  of
comparable strength at, and screened beyond, the millimeter range. 
This has been 
 appreciated in the past in the context of gravitational
axions and light string moduli \cite{axi}, and a similar bound  holds
for  Kaluza-Klein modes.

  Besides mesoscopic gravity experiments, 
one may worry about bounds  from  precision observables  of the
 Standard Model, and  from various  exotic processes. 
Precision  tests of the SM and
compositeness bounds cannot however  rule out in a model-independent way any
new physics above the TeV-ish  scale. 
Bounds for instance from LEP data on
four-fermion operators, or bounds on dimension-five operators
contributing to the  $g-2$ 
of the electron/muon are  safe, as long as  the characteristic scale of
the new physics is a few TeV \cite{ADD}. 
Proton decay and other exotic processes could  of course
rule out large classes of low-scale models, but natural suppression
mechanisms do exist.
One type  of model-independent 
exotic process is  graviton emission  in the bulk, which could be seen
as missing-energy events in  collider experiments.
 The process  is however
suppressed by the four-dimensional Newton constant  at low
 energies,  and only becomes
 appreciable (as one should expect) 
near string  scale  where quantum gravity effects
 are strong \cite{missing} (see also Peskin's talk \cite{peskin}).

   Let me come finally  to question  of the gauge hierarchy.
Although the brane world with a TeV string scale does not really
 solve  the
problem, it  does  transform  it in an interesting way. Instead
of asking why is $M_{\rm Planck}$ so much bigger than
the SM scale, one now asks why is the  transverse size $r$ so
much larger than the string length? In the particular case of
an effectively  two-dimensional transverse space, there is 
furthermore  logarithmic sensitivity on $r$, 
which is (superficially) the  natural setup for obtaining a hierarchy
of scales \cite{AB}.
It is, however,  very hard to know whether any of this 
constitutes progress or not, without fully adressing the issue
of vacuum stability.

\section{Outlook}

 The titles of three out of the last four sections
 had a questionmark in them, and it is often said
that `if you see a question asked in
 a title the answer is NO'.  Well, maybe so, but I am ready to
bet that there {\it is} life without supersymmetry -- we are
after all the living  proofs of it. As for the QCD string, even if
this new attempt to master it fails, the ideas that underly it 
have  already had a profound impact in the field
and are most likely here to stay.
What about the  brane world and TeV strings?
The recent developments   have sunk in
 the point that  we really  know very little about gravity. 
They have also brought string theorists and
phenomenologists closer together, which is  a marvelous thing. 
They have not yet, however, solved any 
theoretical  problem, and abandoning the traditional
hypothesis  does  make gauge coupling
unification -- our one solid empirical hint about physics
beyond the SM  -- look  accidental.  As theorists
we should thus keep an open-minded but  cool head towards the possibility,
which surely deserves further scrutiny.
Ultimately it is  vacuum selection and stability,
or the LHC,  that will of course  decide!

\vskip 1cm

{\bf Aknowledgements}

  I thank Prof. Kummer, Prof.  Roos   and the other members of
the organizing committee  for the invitation to speak, and for
their kind hospitality in Tampere. I also thank the 
New High Energy Theory Center at the University of Rutgers for 
hospitality while writing up this talk.

\end{document}